\documentstyle[12pt]{article}

\begin{document}
\rightline{ILL-(TH)-94-15}
\rightline{hep-ph/9501256}
\rightline{Phys.\ Lett.\ {\bf B340}, 236 (1994)}
\bigskip\bigskip
\begin{center}
{\Large \bf Can the Electroweak Interaction Break Itself?} \\
\bigskip\bigskip\bigskip\bigskip
{\large\bf S.\ Willenbrock} \\
\medskip
Physics Department \\
University of Illinois \\
1110 W. Green Street \\
Urbana, IL\ \ 61801 \\
\end{center}
\bigskip\bigskip\bigskip

\begin{abstract}
I examine the possibility that the electroweak interaction breaks itself
via the condensation of fermions in large representations of the weak SU(2)$_L$
gauge group.
\end{abstract}

\newpage

\addtolength{\baselineskip}{9pt}

\section*{}

\indent\indent The strong and electroweak interactions are described by gauge
theories, based on the gauge groups SU(3)$_c$, SU(2)$_L$, and U(1)$_Y$, acting
on
three generations of quarks and leptons.  At an energy scale $(\sqrt{2}
G_F)^{-1/2} \approx 250$ GeV, the groups SU(2)$_L\times$U(1)$_Y$ are
spontaneously broken to electromagnetism, U(1)$_{EM}$, giving rise to masses
for
three of the electroweak gauge bosons, the $W^+$, $W^-$, and $Z$.  To produce
the
spontaneous breaking of the electroweak symmetry, an additional ingredient must
be added to the theory.  In the standard Higgs model, an SU(2)$_L$ doublet, $Y
= +\frac{1}{2}$ scalar field is introduced, with a self interaction which
forces the field to acquire a non-zero vacuum-expectation value \cite{WS}.
This breaks
the SU(2)$_L$ and U(1)$_Y$ symmetries, leaving the unbroken subgroup
U(1)$_{EM}$
with generator $Q = T_{3L} + Y$.

An alternative mechanism for electroweak symmetry breaking invokes
dynamical symmetry breaking.  In these models, fermion fields which
carry SU(2)$_L\times$U(1)$_Y$ quantum numbers interact via a force which
produces a fermion-bilinear condensate.  This condensate is constructed to be
an SU(2)$_L$ doublet, thereby breaking the electroweak
gauge group to electromagnetism.  An unbroken, global ``custodial''
SU(2) symmetry is sufficient
to ensure that $M_W/M_Z = \cos\theta_W$, a relation that is otherwise not
automatic in models of dynamical electroweak symmetry breaking \cite{TC,SSVZ}.
An example of a model of dynamical electroweak symmetry breaking is
technicolor,
in which a non-Abelian gauge interaction produces a condensate
of technifermions \cite{TC}.

The ordinary strong interaction can dynamically break the electroweak symmetry
at the scale $(\sqrt{2}  G_F)^{1/2} \approx 250$ GeV if it acts on fermions in
larger representations of SU(3)$_c$ than the fundamental representation, such
as color sextets, octets, or decuplets \cite{M,KSS}.  In the one-gluon-exchange
approximation, the force between fermions in complex-conjugate representations
$R$ and $\overline R$ forming a color singlet is proportional to
\begin{equation}
F \sim - \alpha_s C_2(R) \label{ONE}
\end{equation}
where $C_2(R)$ is the quadratic Casimir operator of the representation $R$
($R$ denotes the dimension of the representation).
Since $C_2(R)$ generally increases with $R$, the effective force is
increased between fermions in larger representations of SU(3)$_c$.
Condensation occurs when $\alpha_s(\mu)C_2(R) \sim 1$, which is potentially
consistent with $\mu \approx$ 250 GeV for $R=$ 6, 8, or 10.

In this paper I explore the possibility that the electroweak interaction is
broken by {\it itself} via the condensation of fermions in large
representations of SU(2)$_L$. The main attraction of this approach is that
it potentially requires no new forces beyond the known electroweak interaction.
An obstacle to the realization of this scenario is the loss of
asymptotic freedom of the SU(2)$_L$ gauge coupling. However, I will argue
that this obstacle may be surmountable.

Gauge theories break themselves if the fermions
condense to a non-singlet representation of the gauge group \cite{RDS,P}.
The generalization of Eq.~(\ref{ONE}) to fermions in the representations $R_1$
and $R_2$ forming a state in the representation $R_c$ is
\begin{equation}
F \sim - \frac{\alpha}{2} [C_2(R_1) + C_2(R_2) - C_2(R_c)] \label{FORCE}
\end{equation}
for a gauge group with coupling $\alpha$.

To explore the possibilities, let's construct a simple model of electroweak
self-breaking.
Consider a left-handed Weyl fermion, $\psi$, in the
representation $R$ of SU(2)$_L$, and a pair of left-handed Weyl fermions,
$\chi_a$ ($a=1,2$), in the $R-1$ representation of SU(2)$_L$.\footnote{$R$
denotes the dimension of the representation; the corresponding
weak ``spin'' is $T_L=(R-1)/2$.} Cancellation of
gauge anomalies demands that $\psi$ have $Y=0$, and the $\chi_a$ have equal and
opposite hypercharges. The global
symmetries of the massless theory, in the limit of zero hypercharge
coupling, are
\begin{equation}
{\rm SU(2)}_L\times{\rm SU(2)}_\chi\times{\rm U(1)}
\end{equation}
where SU(2)$_L$ is the gauge symmetry and SU(2)$_\chi$ is the
flavor symmetry of the $\chi_a$ fields (the global U(1) symmetry will be
discussed later). Write the representation $R$ as
$\psi^{i\cdots j}_\alpha$ with $R-1$
totally symmetric SU(2)$_L$ (upper) indices, and one (lower) Lorentz index.
Similarly, write the pair of $R-1$ representations as
$\chi^{i\cdots j}_{a,\alpha}$ with $R-2$ totally symmetric SU(2)$_L$ indices.
The SU(2)$_L$-doublet, Lorentz-scalar condensate which breaks the
electroweak interaction to electromagnetism is
\begin{equation}
<\psi^{ij\cdots k}_\alpha \chi^{l\cdots m}_{a,\beta}
\epsilon^{\alpha\beta} \epsilon_{jl} \cdots \epsilon_{km}> =
\mu^3\delta^i_a \label{DOUBLET}
\end{equation}
where $\epsilon$ is the 2$\times$2 antisymmetric tensor and $\mu$ is of order
250 GeV.  This condensate
respects the diagonal subgroup of SU(2)$_L\times$SU(2)$_\chi$, which serves as
the custodial SU(2). The fields $\chi_a$ must have hypercharge $\pm
\frac{1}{2}$ to maintain $Q=T_{3L}+Y$.  The fields and their
SU(2)$_L\times$U(1)$_Y$ quantum numbers are summarized in Table 1.

\begin{table}[hbt]
\caption[fake]{SU(2)$_L \times$U(1)$_Y$ quantum numbers of the
left-handed Weyl-fermion fields in a model of electroweak self-breaking.}
\bigskip
\begin{center}
\begin{tabular}{clc}
&\underline{SU(2)$_L$} & \underline{U(1)$_Y$} \\
\\
$\psi$ & $R$ & 0 \\
$\chi_1$ & $R-1$ & $+\frac{1}{2}$ \\
$\chi_2$ & $R-1$ & $-\frac{1}{2}$ \\
\end{tabular}
\end{center}
\end{table}

The quadratic Casimir operator of the
representation $R$ is $C_2(R) = \frac{1}{4}(R^2 - 1)$.  The coefficients of the
effective force between the fermions, Eq.~(\ref{FORCE}), in various channels
are listed in Table 2.
The condensate $R\times R \to 1$ is the most-attractive channel and, for
$R\geq 4$,
$R \times R \to 3$ is the next-most-attractive channel. Thus our first
obstacle is the fact that the desired condensate, Eq.~(\ref{DOUBLET}),
corresponding to $R \times (R-1) \to 2$, is not the most-attractive channel.

\begin{table}[hbt]
\caption[fake]{Coefficient of the effective force between fermions in the
representations $R_1$ and $R_2$ of SU(2)$_L$ forming a state in the
representation $R_c$, in the one-gauge-boson-exchange approximation.}
\bigskip
\begin{center}
\begin{tabular}{rl}
$\underline{R_1\times R_2 \to R_c}$ & $\underline{C_2(R_1)+C_2(R_2)-C_2(R_c)}$
\\
\\
$R\times R \to 1$ & $\frac{1}{2}(R^2-1)$ \\
$R\times R \to 3$ & $\frac{1}{2}(R^2-1)-2$ \\
$R\times (R-1) \to 2$ & $\frac{1}{2}(R^2-1)-\frac{1}{2}(R+1)$ \\
$(R-1)\times (R-1) \to 1$ & $\frac{1}{2}(R^2-1)-(R-\frac{1}{2})$ \\
\end{tabular}
\end{center}
\end{table}

Before we confront this obstacle, let's estimate the dimensions of the
representations needed for the electroweak interaction to break itself.  For
the channel $R\times (R-1) \to 2$, the effective weak force is (see
Eq.~(\ref{FORCE}) and Table 2)
\begin{equation}
F \sim - \frac{\alpha_W}{4} [R^2-R-2]
\end{equation}
where $\alpha_W \equiv \alpha(M_Z)/\sin^2\theta_W \approx 1/30$.  The
critical coupling for the channel $R\times R \to 1$ to condense is calculated
in
Ref.~\cite{KSWSSS} to be $C_2(R)\alpha_W \approx 0.3$.  Thus we expect
the channel $R\times (R-1) \to 2$ to condense when
\begin{equation}
\alpha_W \frac{1}{4} (R^2 - R - 2) \geq 0.3
\end{equation}
which implies $R \geq 7$, that is, $T_L \geq 3$.

In the case of odd $R$, fermion
masses are allowed by the SU(2)$_L\times$U(1)$_Y$ gauge symmetries.
The $\psi$ field has a mass term of the form
\begin{equation}
\psi^{i\cdots j}_\alpha \psi^{k\cdots l}_\beta \epsilon^{\alpha\beta}
\epsilon_{ik} \cdots \epsilon_{jl} \label{PSI}
\end{equation}
and the $\chi_a$ fields have a mass term
\begin{equation}
\chi^{i\cdots j}_{a,\alpha} \chi^{k\cdots l}_{b,\beta} \epsilon^{ab}
\epsilon^{\alpha\beta} \epsilon_{ik} \cdots \epsilon_{jl}
\end{equation}
which respects SU(2)$_\chi$. Thus the SU(2)$_L\times$U(1)$_Y$ and SU(2)$_\chi$
symmetries are vector-like.  Since vector symmetries are preserved in
vector-like
gauge theories \cite{VW}, the weak interaction cannot break itself if $R$ is
odd.

For even $R$, a mass term for the
$\psi$ field is forbidden by the SU(2)$_L$
gauge symmetry.  This term would be of the form Eq.~(\ref{PSI}),
which vanishes for even $R$, keeping in mind that the fermion fields
anticommute. A mass term is allowed for the $\chi_a$ fields of the form
\begin{equation}
\chi^{i\cdots j}_{1,\alpha} \chi^{k\cdots l}_{2,\beta} \epsilon^{\alpha\beta}
\epsilon_{ik} \cdots \epsilon_{jl} \label{CHI}
\end{equation}
which violates the global SU(2)$_\chi$ symmetry and thus violates the
custodial SU(2).  We can forbid this term by {\it requiring} global
SU(2)$_\chi$ symmetry (in the limit of vanishing hypercharge), promoting it
from
an ``accidental'' symmetry to a
fundamental one, perhaps left over from a previous stage of symmetry
breaking.

If we assume that fermion condensation does not break Lorentz invariance,
then a condensate in the channel $R\times R \to 1$ cannot form, since it is
proportional to the
expectation value of Eq.~(\ref{PSI}), which vanishes. For $R\geq 4$, the
most-attractive Lorentz-scalar channel is $R\times R \to 3$,
corresponding to the condensate
\begin{equation}
\phi^{ij} = <\psi^{ik\cdots l}_\alpha
\psi^{jm\cdots n}_\beta \epsilon^{\alpha\beta}
\epsilon_{km}\cdots\epsilon_{ln}>\;. \label{TRIPLET}
\end{equation}
This condensate is an SU(2)$_L$ triplet, and thus does not yield
$M_W/M_Z = \cos \theta_W$.  Since it is more attractive than the desired
condensate, Eq.~(\ref{DOUBLET}), it is difficult to see how it could be
sufficiently suppressed not to conflict with phenomenology.

Let us nevertheless assume that this condensate is suppressed, and that
the desired condensate, Eq.~(\ref{DOUBLET}), breaks the electroweak
symmetry. In the
limit of vanishing hypercharge coupling and vanishing mass term for the
$\chi_a$ fields, Eq.~(\ref{CHI}), the model has an
SU(2)$_L$-anomaly-free global U(1) symmetry, with the $\chi_a$ fields of unit
charge and $\psi$ of charge $-2(R-2)/(R+1)$.  This symmetry is broken by
the condensate Eq.~(\ref{DOUBLET}), yielding a massless Goldstone boson.
This symmetry has a U(1)$_Y$ anomaly, but since there are no Abelian
instantons, this Goldstone boson remains massless.

The spontaneous symmetry breaking also leaves massless fermions.  Under the
unbroken custodial SU(2) symmetry, the fermions $\psi$ and $\chi_a$
decompose as shown below:
\begin{equation}
\begin{array}{crcl}
&\underline{{\rm SU(2)}_L\times {\rm SU(2)}_\chi} & \to &
\underline{{\rm SU(2)}} \\
\\
\psi & R\times 1 & \to & R \\
\chi_a & (R-1)\times 2 & \to & R+(R-2) \\
\end{array}
\end{equation}
The custodial-SU(2) $R$ representations pair up to produce massive Dirac
fermions, but there are $R-2$ massless $\chi$ fields left over from the
condensation. When hypercharge is turned on the model has no unbroken
global symmetries (besides U(1)$_{EM}$), so these fermions presumably
obtain a small mass,
but they are potentially too light to be phenomenologically acceptable.

The Goldstone boson and the light fermions can be avoided by explicitly
breaking the chiral symmetries from which they arise, for example by coupling
the fermions to additional gauge fields, or by introducing a mass term
for the $\chi_a$ fields, Eq.~(\ref{CHI}).  However, the
light-fermion masses are protected by the custodial SU(2), and there is a
limitation on the amount of explicit custodial-SU(2) breaking which is
phenomenologically acceptable.

A truly realistic theory must also account for the generation of quark and
lepton masses.  This is where models based on dynamical electroweak symmetry
breaking may run into phenomenological difficulties.  It is not clear how
the present approach could help alleviate those problems.

Another obstacle to electroweak self-breaking is the loss of asymptotic
freedom of the
SU(2)$_L$ interaction in the presence of fermions in large representations.
The one-loop SU(2)$_L$ beta-function coefficient for three
generations of quarks and leptons, a Weyl fermion in the $R$
representation and two in the $(R-1)$ representation,
is\footnote{The Dynkin index for the $R$ representation of
SU(2)$_L$ is $T(R)=\frac{1}{12}R(R^2-1)$.}
\begin{equation}
b_2 = \frac{1}{(4\pi)^2} \left[-\frac{10}{3}+\frac{1}{6}R(R-1)^2\right]
\end{equation}
and asymptotic freedom is lost for $R\geq 4$.  Without asymptotic freedom, it
is difficult to understand why the condensate forms at the electroweak
scale, and not at some higher scale, where the coupling is even larger.

A possible route around this obstacle is to hypothesize that a gauge theory
with fermions in large representations possesses a non-trivial ultraviolet
fixed point.  A similar hypothesis has been studied for QED
\cite{FK,FGMS,LLB,KDK,H,GHRSS,ADG} and for a non-Abelian gauge theory
with a large number of fermions \cite{BZ,KS,MP,GR}. The one-loop beta function
is positive, but
for fermions in large representations perturbation theory may be unreliable,
even for $\alpha_W \approx \frac{1}{30}$.  The ratio of the beta-function
coefficents for the $(n+1)^{th}$ and $n^{th}$ loops $(n \geq 2)$ is
proportional to $T(R)\alpha_W/4\pi$ in the large-$R$ limit ($T(R)$ is the
Dynkin
index of the $R$ representation\footnotemark[1]), and can be of
order unity for sufficiently large $R$. For a Weyl fermion in the
representation $R$ and two in the representation $R-1$, the ratio of the
three-loop and two-loop
beta-function coefficients in the large-$R$ limit (in the MS scheme) is
\cite{TVZ}
\begin{equation}
-\frac{11}{72}R^3\frac{\alpha_W}{4\pi}
\end{equation}
which is approximately of unit magnitude for $R=13$.  For $R=8$, the smallest
acceptable value, this ratio is
about $-0.21$, which suggests that perturbation theory is valid.  Additional
large-$R$ representations may be necessary in this case to produce the
desired breakdown of perturbation theory.

The model considered in this paper is free of gauge anomalies.  However, one
must also consider the discrete SU(2)$_L$ anomaly \cite{W}.
This anomaly vanishes if the sum of the Dynkin indices of the left-handed
fermions is an integer.  For the above model,
\begin{equation}
\sum_r T(r) = \frac{1}{4}R(R-1)^2
\end{equation}
where the sum is over the representations. This is half-integer only
for $R=4N+2$, where $N$ is an integer. Thus the case of $R=4N$ is free
of anomalies, which includes $R=8$.

In this paper I have suggested that the electroweak symmetry can break itself
via the condensation of fermions in the $R$ and $R-1$ representations of
SU(2)$_L$, for $R$ even and $\geq 8$.
I hope this will provide motivation for the investigation of the
non-perturbative behaviour of non-Abelian
gauge theories with fermions in large representations.
If the answer to the question posed in the title should prove to be
affirmative, the next question will be: {\it Does} the electroweak interaction
break itself?

\section*{Acknowledgements}

I am grateful for conversations with W. Bardeen, E. Eichten, A. El-Khadra,
H. Georgi, A. Koci\'c, J. Kogut, S. Raby, and J. Shigemitsu. I am grateful for
the hospitality of the high-energy theory groups at Ohio State University
and Fermilab, where part of this work was performed.

\clearpage

\end{document}